\documentstyle[prl,aps,preprint]{revtex}

\begin{document}
\draft \title {Spatial Correlation in Quantum Chaotic Systems with
  Time-reversal Symmetry: Theory and Experiment}

\author{V. N.  Prigodin$^1$\cite{address-r}, Nobuhiko
  Taniguchi$^2$\cite{address-j}, A.  Kudrolli$^3$, V. Kidambi$^3$, and S.
  Sridhar$^3$}

\address{$^1$Max-Planck-Institute f\"ur Physik komplexer Systeme, Au\ss
  enstelle Stuttgart, Heisenbergstr.1, 70569 Stuttgart, Germany\\
  $^2$Department of Physics, Massachusetts Institute of Technology,
  Cambridge, MA 02139\\$^3$Department of Physics, Northeastern University,
  Boston, MA 02115}

\maketitle
\begin{abstract}
  The correlation between the values of wavefunctions at two different
  spatial points is examined for chaotic systems with time-reversal
  symmetry.  Employing a supermatrix method, we find that there exist
  long-range Friedel oscillations of the wave function density for a given
  eigenstate, although the background wavefunction density fluctuates
  strongly. We show that for large fluctuations, once the value of the
  wave function at one point is known, its spatial dependence becomes
  highly predictable for increasingly large space around this point.
  These results are compared with the experimental wave functions obtained
  from billiard-shaped microwave cavities and very good agreement is
  demonstrated.
\end{abstract}


\narrowtext

Quantum properties of classically chaotic systems such as ``billiards'' and
quantum dots are revealed to have remarkable universal behaviors which depend
only on the generic symmetry of the system, such as the time-reversal symmetry
and/or the spin rotational symmetry~\cite{Berry91,Haake,alt1}.  It has been
shown that the spectral statistics are well described by universal statistical
correlations derived from the random matrix theory~\cite{Wigner51,Dyson62}.
(See also \cite{Mehta,Bohigas91,Heller93} for reviews.)  Complementary and
comprehensive information beyond the energy statistics can be obtained by
examining the statistics of chaotic wavefunctions.  For example, the
distribution of the local density in a fully chaotic system is known to be
universal and obey the Porter-Thomas distribution~\cite{Berry91,PorterThomas},
which is given for a system with time-reversal symmetry, by the equation
\begin{equation}
  P_0(v) \equiv \langle \delta(v- V|\psi_{\epsilon}({\bf r})|^2)\rangle
={1\over\sqrt{2\pi v}}\;\exp(-v/2),
\label{PorterThomas}
\end{equation}
where $\psi_{\epsilon}({\bf r})$ is the eigenfunction with energy
$\epsilon$ in a system with volume $V$. $\langle \cdots\rangle$ means the
average over the disorder and/or irregular potential.
Eq.~(\ref{PorterThomas}) tells us that wavefunctions fluctuate strongly
but in a universal way.

To get further understanding about the nature of chaotic wavefunctions, other
statistical quantities which can characterize their spatial correlations are
desirable and needed.  The average behavior of the amplitude of the
wavefunction has been conjectured by Berry to be similar to a pattern
generated from random super-position of plane waves \cite{berry81,oconnor87}.
Based on this assumption the average amplitude correlations were shown to be a
Bessel function.  Recently, these expressions have been derived within the
supersymmetry formalism \cite{Prigodin94}.  In this paper, taking a chaotic
system with the time-reversal symmetry, correlations about a particular value
of the wavefunction have been derived analytically and compared to experiments
for the first time. This is not only a more stringent test of the universality
of chaotic wavefunctions, but also gives us a handle on knowing the behavior
of a wavefunction of the system, once the wavefunction is known only at a
limited number of points. Our main object is to investigate the joint
probability distribution function of the density for two different spatial
points ($r=|{\bf r}_1 -{\bf r}_2|$) defined by
\begin{eqnarray}
&&P(v_1,v_2;r) = \left\langle\delta (v_1 - V|\psi _{\epsilon}({\bf
    r}_1)|^2)\:\delta (v_2 - V|\psi _{\epsilon}({\bf
    r}_2)|^2)\right\rangle.\nonumber\\
\label{Def:Pv1v2}
\end{eqnarray}
Although the relevant universality class for experiments which can
directly observe the amplitude of wavefunctions, such as quantum
corrals~\cite{Crommie93,Hasegawa93} and microwave
cavity~\cite{microwave,sridhar91} is orthogonal, only the expression for
$P(v_1,v_2)$ in the unitary case is known so far~\cite{Prigodin94b},
because of technical difficulties.  Here we evaluate $P(v_1,v_2;r)$ for
the orthogonal case by finding special techniques (Eq.~(\ref{Q-relation})
below).  In the microwave cavity, the electromagnetic field obeys the same
equation of motion as a quantum particle in a two-dimensional billiard.
This enables us to make direct comparison between the analytical results
and the experimental data.  The experimental data for the wave function
density was obtained from thin cylindrical microwave cavities of the Sinai
stadium, by using a cavity perturbation
technique~\cite{sridhar91,microwave1}. The wavefunction density data was
earlier seen to be consistent with
Eq.~(\ref{PorterThomas})~\cite{microwave1}, and also in agreement with the
expression for density auto-correlations obtained in
ref.~\cite{Prigodin94}.  Once we know $P(v_1,v_2;r)$, we can also find the
conditional probability
\begin{eqnarray}
P_{v_1}(v_2;r) = P(v_1,v_2;r)/P_0(v_1)\:.
\label{jointPv1v2}
\end{eqnarray}
$P_{v_1}(v_2;r)$ describes the distribution of the wave function of $v_2=V
|\psi_{\epsilon}({\bf r}_2)|^2$ at the point ${\bf r}_2$, provided that $v_1=V
|\psi_{\epsilon}({\bf r}_1)|^2$ at ${\bf r}_1$.  Here we compare the coordinate
dependence of the first and the second moments of Eq.~(\ref{jointPv1v2})
between theory and experiment.

The system under consideration can be expressed by the Hamiltonian
\begin{equation}
  H = {1 \over 2m} {\bf p}^2 + U_0({\bf r}) + U_1({\bf r}),
\label{hamiltonian}
\end{equation}
where $U_0({\bf r})$ denotes the regular part of a confining potential, and
$U_1({\bf r})$ is a random potential which is responsible for the chaotic
dynamics --- impurities or `imperfection' of the shape of the system.  We take
the ensemble average over $U_1({\bf r})$ by use of the supermatrix method,
which reproduces the spectral correlations of Wigner-Dyson
statistics~\cite{Efetov83,Verbaarschot85b} and recently was successfully
applied to calculate other universal properties relating to chaotic
wavefunctions~\cite{Prigodin94,Prigodin94b,Taniguchi94b,Falko94}.

We should remark that although the supermatrix method was originally derived
from the Gaussian random potential where the mean free path $\ell$ is much
smaller than the size of the system $L$, the ergodicity
hypothesis~\cite{Lee85} allows us to extend our present result for cases
$\ell\sim L$ by identifying the averaging over space and different states for
a given sample with that over disorder. As a confirmation of the ergodicity
hypothesis we will demonstrate that the theoretical dependencies for
wavefunctions derived from a disordered system with $\ell \ll L$ are universal
and describe very well experimental results for quantum billiard systems for
which $\ell \sim L$.

After lengthy calculations which we will sketch later, we have obtained
the following analytical expression for $P(v_1,v_2;r)$ for the orthogonal
case:
\widetext
\begin{eqnarray}
P(v_1,v_2;r) = {1 \over 2\pi f(r)}\:\int^{f(r)}_0{pdp \over
    \sqrt{f^2(r)-p^2}}\:\left(1 + p~{d \over dp}\right)\:\sqrt{{1-p^2 \over
      2\pi v_1v_2}}\:\int^{\infty}_0{dz
    \over \sqrt {z}}\:e^{z/2}~\varphi({v_1+z
    \over 2},{v_2+z \over 2};p),
\label{result:Pv1v2}
\end{eqnarray}
\narrowtext
\begin{eqnarray}
\varphi(v_1,v_2;p) = {1\over 1-p^2}
I_0\left({2p\sqrt{v_1v_2}\over 1-p^2}\right)\exp{\left(-{v_1+v_2 \over
    1-p^2}\right)},
\label{def:varphi}
\end{eqnarray}
\noindent where $I_0(p)$ is the modified Bessel function and $f(r)$ is the
Friedel function \cite{Friedel}. Note that $P(v_1,v_2;r)$ depends
on $f(r)$ in an universal way and all coordinate dependence of $P(v_1,v_2;r)$
is incorporated only through $f(r)$.  In fact, the function $f(r)$ represents
the average correlation of the amplitude of wave function
\cite{berry81,oconnor87}: $f(r) = V^2<
\psi^*_{\epsilon}(r_1)\psi_{\epsilon}(r_2)>$.  For the case of a flat
background potential in a $d$-dimensional system, $f(r)$ becomes
\begin{equation}
f(r)=\Gamma(d/2)\:(2/kr)^{d/2-1}~J_{d/2-1}(kr)\:e^{-r/2\ell},
\end{equation}
where $k$ is the wave vector ($\epsilon =\hbar^2 k^2/2m$), $J_n(x)$ is the
Bessel function, and $\Gamma(n)$ is the gamma function.  Note that the
envelope of $f(r)$ decays like $(kr)^{-(d-1)/2}$ for $k^{-1}\lesssim
r\lesssim\ell$ and this behavior corresponds to the representation of chaotic
wave function as a random superposition of plain waves
\cite{berry81,oconnor87}.

It should be remarked that the same distribution function in the unitary
case is given by $P(v_1,v_2;r)=\varphi (v_1,v_2;f(r))$~\cite{Prigodin94b}.
Due to two additional integrations in Eq.~(\ref{result:Pv1v2}) the
spatial correlations for the orthogonal case are weaker and the
fluctuations are stronger in comparison with the unitary one.

We first check that $P(v_1,v_2;r)$ given by Eq.~(\ref{result:Pv1v2}),
yields correct limiting behaviors.  For remotely separate points such
that $f(r)\approx 0$ the fluctuations of the wavefunction density within
the given
eigenstate become independent, i.e., $P(v_1,v_2;r) \approx P_0(v_1)
P_0(v_2)~$.  In the opposite limit of close enough points that $f(r)\approx
1$, there is an obvious strong correlation between fluctuations as
\begin{eqnarray}
P(v_1,v_2;r) \approx {P_0(v_1) \over \sqrt {8\pi v_1(1-f^2)}} \exp
\left[ -{(v_1 -v_2)^2 \over 8v_1(1-f^2)} \right]\:.
\label{limit}
\end{eqnarray}

{}From Eq.~(\ref{limit}), we can extract information about the gradient of
the wavefunction.  By setting $\psi({\bf r}_2)=\psi({\bf r}_1)+r\nabla_{\bf
  n}\psi({\bf r}_1)+O(r^2)$, and expanding for small $r$, we obtain the joint
distribution involving the wavefunction and its gradient in any direction
${\bf n}={\bf r}/r$.  Accordingly we find that the gradient of the
wavefunction along any direction fluctuates independently of the value of
wavefunction, and obeys also the Porter-Thomas distribution:
\begin{equation}
  \langle \delta(v-V|\psi({\bf r})|^2) \delta(s-{Vd\over
    2k^2}|\nabla_{\bf n}\psi({\bf r})|^2) \rangle = P_0(v) P_0(s).
\end{equation}
This conclusion is, however, not true for higher gradients of the
wavefunction.

The conditional probability $P_{v_1}(v_2;r)$ is obtained straightforwardly
with Eqs.~(\ref{jointPv1v2},\ref{result:Pv1v2}).  To see how the
fluctuations of the wavefunctions behave and to compare between the
analytical results and the experiments, the conditional average $\langle
v_2\rangle_{v_1}$ and the conditional variance is more convenient.
Denoting $\delta v_2=v_2-\langle v_2 \rangle_{v1}$, we obtain:
\begin{eqnarray}
  \langle v_2\rangle_{v_1}&=& 1 + f^2(r)\;(v_1-1),\label{average}\\
  \langle(\delta v_2)^2\rangle_{v_1} &=& 2 + 4f^2(r)\;(v_1 - 1) +
  2f^4(r)\;(1 - 2v_1).
\label{variance}
\end{eqnarray}
Comparing with results obtained for unitary case~\cite{Prigodin94b}, we
find that the conditional average Eq.~(\ref{average}) is exactly the same.
Thus we cannot tell the symmetry of the system only from the
averaged amplitude even if we know the conditional one.  To detect the
symmetry, we have to examine the variance, where there is a factor of 2
difference between the orthogonal and the unitary case.

In Fig.~\ref{fig:average} and \ref{fig:variance}, we compare the analytical
results of the conditional average and variance with the experimental data
from microwave cavities.  The experimental curves were obtained by picking
points in a wavefunction with the same value and calculating the average
wavefunction value a distance $r$ from it on a circle. This quantity was then
again averaged over at least 50 wavefunctions after rescaling the wave number
to obtain better statistics.  Fig.~\ref{fig:average} shows the plot of
$\langle v_2\rangle_{v_1}$ in Eq.~(\ref{average}) with experimental results
for $v_1 = 2$ and $v_1 = 7$.  Very good agreement is seen for both sets.  In
fact, agreement is excellent for all values of $v_1$ above 1, below which the
noise and  errors in the measurement of the
wavefunction measurements lead to qualitative differences.
In Fig.~\ref{fig:variance}, the comparison of the data to the expression
in Eq.~(\ref{variance}) was done.  Again one sees an excellent agreement
with experimental errors of 5 \%, which is the level of experimental accuracy.

According to Eqs.~(\ref{average},\ref{variance}), we can say, as in the
unitary case~\cite{Prigodin94b}, that large fluctuations of the
wavefunction have some striking structure which is not present for small
fluctuations. For $v_1 \gg 1$ the ratio of the variance to the average
square is
\begin{equation}
  {\langle (\delta v_2)^2\rangle_{v_1}\over\langle
    v_2\rangle^2_{v_1}}\approx 2(1-f^2)\;{1+2v_1 f^2\over(1+v_1 f^2)^2}\:.
\end{equation}
Therefore at $r\lesssim \xi$, where the ``correlation length''
$\xi\sim k^{-1} v_1^{1/(d-1)}\gg k^{-1}$, the variance $\langle (\delta
v_2)^2\rangle_{v_1}$ can be very small in comparison with $\langle
v_2\rangle^2_{v_1}$.  It means, that once we know that the wavefunction
$|\psi({\bf r})|^2$ is equal to $v_1$ at ${\bf r}_1$, it is highly likely
to have a value $\langle
v_2\rangle_{v_1}\sim f^2 v_1$ at ${\bf r}_2$ for $r\lesssim \xi$.  In this
sense, the large fluctuation behavior of the wavefunction becomes highly
predictable.  In contrast, for small fluctuations $v_2 \ll 1$, we easily see
that $\langle (\delta v_2)^2\rangle_{v_1} \approx 2\langle v_2
\rangle^2_{v_1}$, independent of $v_1$.  We also find directly from
Eq.~(\ref{result:Pv1v2}) that fluctuations turn out to be independent, i.e.,
$P_{v_1}(v_2;r) \approx 1/\sqrt{v_2}$.  Although more careful evaluation gives
us a correlation length $\xi$ which ensures independent fluctuations for the
region $r\gtrsim \xi$, $\xi$ in such an evaluation turns out to be very
small, i.e., $\xi\sim k^{-1}\sqrt{v_1}\ll k^{-1}$.

These behaviors are qualitatively the same both in systems with or without
time-reversal symmetry.  Therefore we can say that this is a generic property
of chaotic wavefunctions. Also, in the semiclassical description of chaotic
systems, periodic and closed orbits are known to be associated with large
values of the wavefunctions $|\psi({\bf r})|^2$~\cite{heller84,bogomolny}. In
this respect, our present results may imply there is some structure present in
these orbits.

Now let us proceed with the derivation of our main result in
Eqs.~(\ref{result:Pv1v2},\ref{def:varphi}).  To evaluate the joint
distribution $P(v_1,v_2;r)$, we work with its moments,
\begin{eqnarray}
\label{Def:qnm}
q_{nm}(r) = V^{n+m}\:\left\langle |\psi _{\epsilon}({\bf r}_1)|^{2n} |\psi
_{\epsilon}({\bf r}_2)|^{2m}\right\rangle\equiv\langle v_1^n v_2^m\rangle.
\end{eqnarray}
$q_{nm}$ is known to be closely related to the moments of the
{\em exact\/} retarded and advanced Green functions $G^{R,A}_\gamma$,
\begin{eqnarray}
 && F_{nm}(r;\gamma) = {i^{n-m} \over (\pi\nu)^{n+m}}
  \left\langle\left(G^R_\gamma({\bf r}_1,{\bf r}_1)\right)^n
  \left(G^A_\gamma({\bf r}_2,{\bf r}_2)\right)^m\right\rangle~,\nonumber\\
\label{Def:Fnm}
\end{eqnarray}
where $\nu$ is the average DOS and
\begin{eqnarray}
G^{R,A}_\gamma({\bf r}_1,{\bf r}_2) = \sum _\alpha
{\psi _{\alpha}({\bf r}_1)\psi _{\alpha}^*({\bf r}_2) \over \epsilon -
\epsilon_\alpha \pm i\gamma/2}.
\end{eqnarray}
We can obtain $q_{nm}$ in terms of $F_{nm}$ by the
relation~\cite{Wegner80,Altshuler89b}
\begin{equation}
  q_{nm}={(n-1)!\;(m-1)!\over 2\: (n+m-2)!}\lim_{\gamma \to
    0}{\left(\gamma\over \Delta\right)}^{n+m-1}F_{nm}\;,
\label{qnm-Fnm}
\end{equation}
since the leading contribution to $F_{nm}$ for small $\gamma$ comes from
the state whose energy $\epsilon_{\alpha}$ coincides with
$\epsilon$.

$F_{nm}$ can be evaluated by the supermatrix method.  However,
since we cannot utilize a simple expression like Eq.~(12) of
Ref.~\cite{Prigodin94b} for the orthogonal case, we are forced to
expand $F_{nm}$ directly by the Friedel function $f(r)$ as (see also
Ref.~\cite{Prigodin94,Taniguchi94b})
\begin{equation}
  F_{nm}(r;\gamma)= n!\, m!\sum_q~C_q(n,m)\,f^{2q}(r).
\label{Fnm-expansion}
\end{equation}
Using the same notation for
the supermatrix elements as in Ref.~\cite{Efetov83}, and defining
$\tilde{Q}^{ab}\equiv(-1)^a Q^{ab}$ (for $a=1,2$), the coefficient
$C_q(n,m)$ in Eq.~(\ref{Fnm-expansion}) is given in terms of
$Q$-supermatrix elements:
\begin{eqnarray}
  &&C_q(n,m)= \sum\prod_{a,b=1}^2
  {\left\langle(\tilde{Q}_{34}^{ab})^{k_{ab}}(\tilde{Q}_{43}^{ab})^{p_{ab}}
    (\tilde{Q}_{33}^{ab})^{l_{ab}}\right\rangle_Q \over
    (1+\delta_{ab})^{k_{ab}+p_{ab}}\; k_{ab}!\; p_{ab}!\; l_{ab}!}
\label{Cnm:expansion}
\end{eqnarray}
where the summation is taken over the all possible combinations of nonnegative
integer $k_{ab}$, $p_{ab}$, $l_{ab}$ which satisfy the condition: $2q =
\sum_{a\neq b} (k_{ab}+p_{ab}+l_{ab})$, $m-\sum_a l_{1a}=2k_{11}+\sum_{a\neq
  b}k_{ab}=2p_{11}+\sum_{a\neq b}p_{ab}$, and $n-\sum_a
l_{a2}=2k_{22}+\sum_{a\neq b}k_{ab}=2p_{22}+\sum_{a\neq b}p_{ab}$.  The symbol
$\langle\cdots\rangle_Q$ denotes an integration over the saddle point
manifold, i.e.,
\begin{equation}
\langle \cdots\rangle_Q \equiv \int\!\! DQ\: (\cdots)\exp[-
(\pi \gamma/4\Delta)\mbox{Str}\Lambda Q],
\end{equation}
where the definitions of $\Lambda$ and $\mbox{Str}$ as well as the
structure of the $Q$ matrix are found in~\cite{Efetov83}.

In principle, averaging $\langle\cdots\rangle_Q$ in Eq.~(\ref{Cnm:expansion})
can be carried out by using the parameterization of Ref.~\cite{Efetov83}.
However, we found it technically unfeasible to evaluate this expression in
such a general form.  Fortunately, to get Eq.~(\ref{qnm-Fnm}) one need only to
know $F_{nm}(\gamma \to 0)$.  The leading contribution in this limit can be
extracted by transforming parameters $\lambda_i=1+u_i \sqrt{\Delta/\gamma}$
for $i=1,2$ in the parameterization given in Ref.~\cite{Efetov83}.  By
calculating the leading order of small $\gamma$, we find the relation
  \begin{eqnarray}
    \tilde{Q}^{ab}_{cd}\;\tilde{Q}^{a'b'}_{c'd'}&\simeq&
    \tilde{Q}^{ab'}_{cd'}\;\tilde{Q}^{ab'}_{cd'},
\label{Q-relation}
\end{eqnarray}
for $a,b=1,2$ and $c,d=3,4$.  Substituting Eq.~(\ref{Q-relation}) into
Eq.~(\ref{Cnm:expansion}) and
combining with Eq.~(\ref{qnm-Fnm}), we finally obtain after integrating
over the $Q$ matrix,
\begin{eqnarray}
  q_{nm}(r) = \sum_q {(2n-1)!!\:(2m-1)!!\:f^{2q}(r) \over
    2^{n+m-2q}\:(n-q)!\:(m-q)!\:(2q)!}\:.
\label{result:qnm}
\end{eqnarray}
Reconstructing $P(v_1,v_2;r)$ from the moments $q_{nm}$ completes the
derivation of our main result Eqs.~(\ref{result:Pv1v2},\ref{def:varphi}).

In conclusion we have presented analytical results for universal
statistical quantities which characterize the coordinate dependence of
chaotic wavefunctions of the system with time-reversal symmetry.  Further we
have demonstrated excellent agreement between the theoretical results and
experimental results of microwave cavities.  The spatial correlations
demonstrate the long-range Friedel oscillations of wavefunction density and the
existence of extended spatial regions of high wavefunction density.

The authors are grateful to B. L. Altshuler and B. D. Simons for their interest
in the work.  This work is partially supported through NSF Grant No. DMR
92-04480 and ONR under Grant No. N00014-92-J-1666.

\widetext
\begin{figure}
%\centerline{\epsfxsize=8cm \epsfbox{average.eps}}
\caption{
The average spatial dependence of chaotic wavefunction squared $\langle
v_2\rangle_{v_1} =V\langle|\psi_{\epsilon}({\bf r}_2)|^2\rangle$ whose
value $V|\psi_{\epsilon}({\bf r})|^2$ at a point ${\bf r}_1$ is known to
be $v_1$ ( $r=|{\bf r}_2 - {\bf r}_1|$, $V$ is the volume, and
$\hbar^2k^2/(2m)=\epsilon$). The theoretical prediction
Eq.~({\protect\ref{average}}) is compared with experiments from the
microwave Sinai stadium cavity for $v_1=7 (\Box$) and $v_1=3$ ($\circ$).
Inset: Representative eigenfunction of the chaotic Sinai stadium.
}\label{fig:average} \end{figure}

\begin{figure}
%\centerline{\epsfxsize=8cm \epsfbox{variance.eps}}
\caption{
Comparison of the conditional variance of a wave function $\langle (\delta
v_2)^2\rangle_{v_1}$ ($\delta v_2 = v_2 - \langle v_2\rangle_{v_1}$ and
$v_2 = V|\psi_{\epsilon}({\bf r}_2)|^2$) as a function of the distance
from point ${\bf r}_1$ and a reference value $v_1$ ($v_1=
V|\psi_{\epsilon}({\bf r}_1)|^2$) between the theory
(Eq.~({\protect\ref{variance}})) and the experiment for $v_1=7$ ($\Box$)
and $v_1=3$ ($\circ$).
}\label{fig:variance} \end{figure}

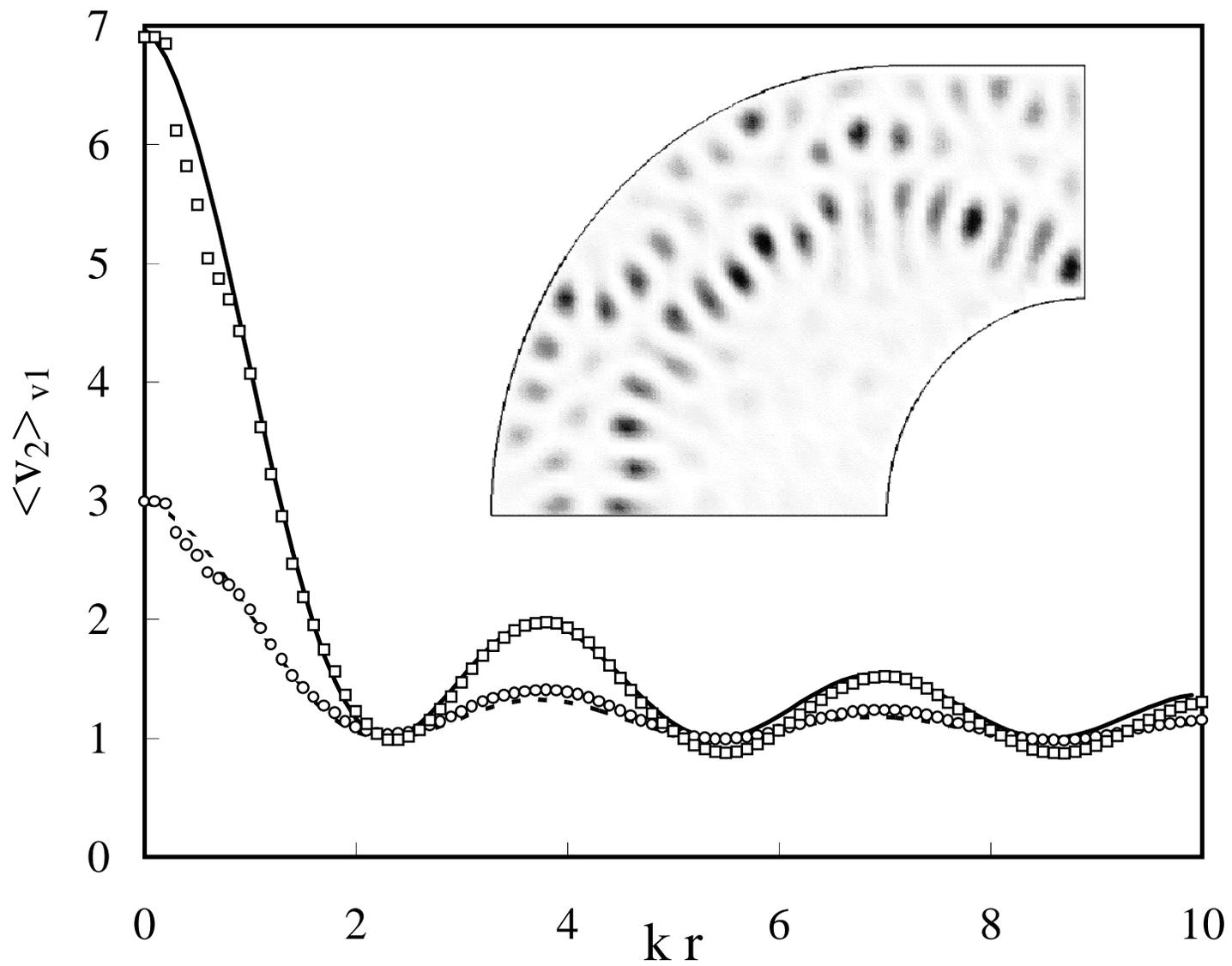
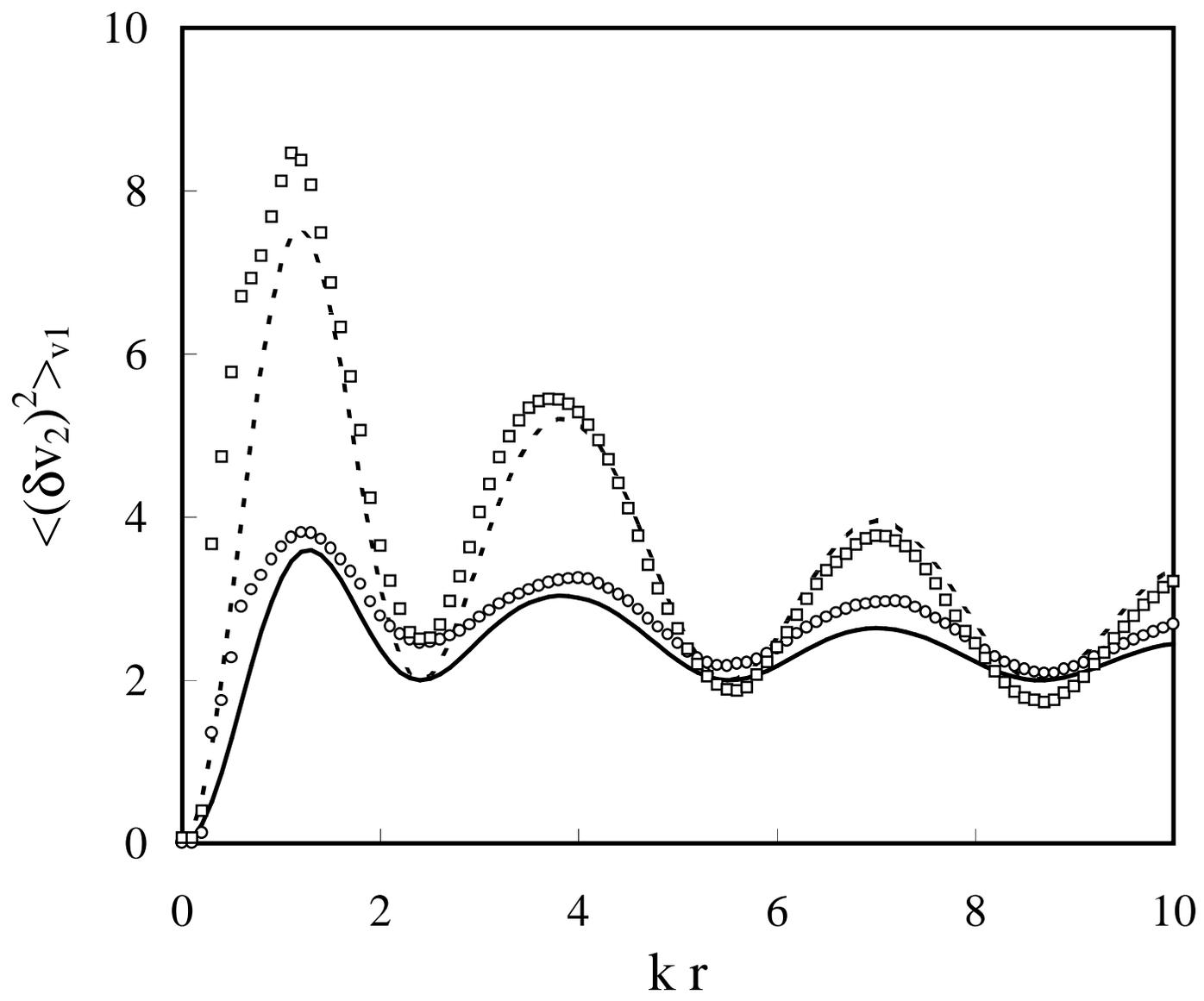
\end{document}